\documentclass[oneside,english]{amsart}
\usepackage[T1]{fontenc}
\usepackage[utf8]{inputenc}
\usepackage{amsthm}
\usepackage{amssymb}

\makeatletter
\numberwithin{equation}{section}
\numberwithin{figure}{section}
\theoremstyle{plain}
\newtheorem{thm}{Theorem}
  \theoremstyle{definition}
  \newtheorem{defn}[thm]{Definition}

\makeatother

\begin{document}

\title{Connection and Dispersion of Computation}

\author{KOBAYASHI, Koji}
\begin{abstract}
In this paper, we describe the impact on the computational complexity
of Connection and Dispersion of CNF. In previous paper \cite{Symmetry and Uncountability of Computation},
we told about structural differences in the P-complete problems and
NP-complete problems. In this paper, we clarify the CNF's dispersion
and HornCNF's connection, and shows the difference between CNFSAT
HornSAT. First we focus on the MUC decision problem. We clarify the
relationship between MUC and the classifying of the truth value assignment.
Next, we clarify the clauses correlation and orthogonal by using the
two inner product of clauses. Because HornMUC has higher orthogonal,
its orthogonal MUC is polynomial size. Because MUC has higher correlation,
its orthogonal MUC is not polynomial size. And, HornMUC whereas only
be a large polynomial is at most its size even if orthogonal than
orthogonal high, MUC will be fit to size polynomial in the size and
orthogonalized using HornCNF more highly correlated shown. So $DP\neq P$,
and $NP\neq P$.
\end{abstract}
\maketitle

\section{CNF's classification and CNFSAT}

We show the relationship between CNFSAT and CNF's classification.
We show the relationship between MUC decision problem and CNFSAT.
And We show the relationship that determined by the CNF. And We show
the relationship between CNF's classification and MUC dicition problem.

\subsection{MUC decision problem}

Describes the MUC decision problem. MUC decision problem is the problem
to decide the CNF is MUC (Minimum Unsatisfiable Core) or not. MUC
is the unsatisfiable CNF. And it changes MUC to satisfiable CNF that
remove one of the MUC's clause. MUC decision problem is combination
problem of coNP-complete and P-complete problem. And HornCNF's MUC
dicision problem is P-complete because of $P=coP$.

The relationship between the DP-complete and P-complete is;
\begin{thm}
\label{thm:Relation between DP and P}If $P\neq DP$ then $P\neq NP$.
So if we can not reduce MUC dicision problem to HornMUC dicision problem
in polynomial time, then $P\neq NP$.\end{thm}
\begin{proof}
If $P=NP$ then $NP=coNP$ and $P=DP$. So MUC dicision problem can
reduce HornMUC dicisition problem in polynomial time. So take the
contrapositive, if we can not reduce MUC dicision problem to HornMUC
dicision problem in polynomial time, then $P\neq NP$.
\end{proof}

\subsection{CNF classification}

Describe the relationship that define the CNF. CNF clauses value corresponds
to either true or false. Clauses are the rules that maps each truth
value assignment to truth value. This is equivalence relation that
classify each truth values to equivalent class.
\begin{defn}
\label{def:CNF equivalence relation}Clauses equivalence relation
is the truth value assignments relation that equal the clauses value.
Similarly, CNF equivalence relation is the truth value assignments
relation that equal the all clauses value set.

I will use the term “CNF classification” to the truth value assignments
equivalence class of CNF, and {}``CNF equivalence class'' to the
equivalence class of CNF classification. And “Logical value assignment”
to the set of the clauses values. And {}``Cyclic value assignment''
to the logical value assignment that have only one false value clause,
and {}``All true assignment'' to the logical value assignment that
have no false value. Number of cyclic value assignment matches the
number of the clauses. All true assignment is only one. The combined
truth value assignment and logical value assignment is the truth value
table of the clauses. Especially, I will use the term {}``Logical
value table'' to the truth value table.
\end{defn}

\subsection{CNF classification and MUC dicision problem}

MUC decision problem is a matter to determine the truth as an input
CNF, can be thought as the problems dealing with CNF classification.
The problem is the decision problem that the logical value assignment
includes all the cyclic value assignment and excludes all tru assignment.
So MUC decision problem can be divided into two calculations, CNF
classification and decision of the logical value assignments.

CNF classification includes the difference of MUC decision problem
and HornMUC decision problem. In decision of logical value assignment,
There is no difference between MUC decision problem and HornMUC decision
problem. And these can be determined in polynomial time either. Especially,
all logical value assignments is cyclic value assignment, the decision
of logical value assignment can determined in polynomial time of the
number of the clauses.
\begin{thm}
\label{thm:1.3 CNF classification and MUC dicision problem}Decision
of logical value assignment can be done in polynomial time either
MUC decision problem or HornMUC decision problem. Thus, the difference
in computational complexity of the MUC decision problem and HornMUC
decision problem will appear in size of the logic value assignments
of CNF classification.\end{thm}
\begin{proof}
The decision of logical value assignment of MUC decision problem and
HornMUC decision problem is the computation that the logical value
assignment includes all cyclic value assignment and excludes all truth
assignment. We can determine the decision by determine the all logical
value assignment. Thus, We can determine the decision only the polynomial
time of the logical value assignment. 

And we can handle in polynomial time of the logical value assignment
that reduce from MUC decision problem and HornMUC decision problem.
So, if they have the coputation complexity difference between MUC
decision problem and HornMUC decision problem, the difference included
in CNF classification.
\end{proof}
Therefore, What has to be noticed is the CNF classification.

\section{MUC as Periodic function}

In view of periodic function, let us then consider MUC. CNF clauses
classification have the periodicity in truth value table. Thus, we
can deal with CNF as periodic function. 

The truth table grows larger the type of variables included in the
clauses, and truth value assignment is changed and the clause is false
by variable's positive and negative. Thus, in periodic function of
clauses, variable is cycle and variable's positive/negative is phase.
And CNF is the periodic function that put many clauses periodic function.
Clauses is the notation in the frequency domain, logical value table
is the notation in the time domain.
\begin{defn}
\label{def:Clause cycle}I will use the term {}``Clause cycle''
to the number of the variables in clauses. The number of Clause cycle
is equal to the number of all cases of change the clauses variable's
positive/negative.

\label{def:Clause phase}I will use the term {}``Clause phase''
to the positive/negative configuration of the variables in clauses.
Clause phase is the position of the truth value assignment that the
clause make false in truth value table. The clause phase difference
of two clauses is equal the minimum Hamming distance of these truth
value assignment that make these clauses false.
\end{defn}
In MUC, there is a equivalence that makes only each clause false,
and the truth value assignment does not belong to the false equivalence
of other clauses. In other words, the equivalence class is not a combination
of other clauses of MUC. Thus, between the clause and the other clauses,
there is the orthogonal that could not save the logical value in transposition.
To put it the other way round, any false clause in the same equivalence
can transpose another clause each other. Thus, between the clause
and the another, there is the correlation that can save the logical
value in transposition.
\begin{thm}
\label{thm:Orthogonal and correlation of MUC}Each clauses of the
MUC have the orthogonality that can not replace with a combination
of other clauses in the MUC. If MUC have the equivalence class that
have some false clauses, the clauses have the correlation that can
replace each other.\end{thm}
\begin{proof}
We prove the clauses orthogonality of MUC using proof by contradiction.
We assume Clauses $C_{1}$ without orthogonality is included in the
MUC. Because $C_{1}$ does not have the orthogonality, the truth value
assignments that $C_{1}$ is false will be false in another clauses.
But this is inconsistent with the terms of the MUC (there are truth
value assignment clause that clause is false). Thus From the proof
by contradiction, clauses of MUC have orthogonality.

It is clear that the clauses have the correlation if the clauses have
the same equivalence class.
\end{proof}
Logic value assignment of MUC and orthogonality/correlation has a
deep relationship. Logical value Assignment represents the relationship
between the some clauses which is false in truth value assignment.
In other words, the same value clauses in same logical value assignment
have correlation. To put it the other way round, the different value
clauses in same logical value assignment have orthogonality. In addition,
there are the cycric value assignment to each clauses in MUC, there
must also be orthogonal of the each clauses. Therefore, to increase
the orthogonality between MUC clauses, it is necessary that all logical
value assignment is cyclic value assignment in MUC.
\begin{thm}
\label{thm:orthogonality and cyclic value assignment}All clauses
in MUC that all logical value assignment is cyclic value assignmet
is orthogonal. And any truth value assignment is false at only one
clause.\end{thm}
\begin{proof}
It is clear from the definition of the cyclic value assignment. 
\end{proof}
In addition, MUC is the CNF that is false of all truth value assignment,
and there is always a clause corresponding to the equivalence classes.
In other words, MUC is complete system based on the equivalence classes
of the truth value assignment.
\begin{thm}
\label{thm:MUC as Completeness}MUC is complete system based on the
equivalence classes of the truth value assignment.\end{thm}
\begin{proof}
It is clear from the definition of the equivalence classes of the
truth value assignment.
\end{proof}
Thus, MUC thats all logical value assignments are cyclic value assignments
is complete orthogonal function.
\begin{thm}
\label{thm:Orthogonal MUC}MUC thats all logical value assignments
are cyclic value assignments is complete orthogonal function. And
every clauses are orthogonal basis of the complete orthogonal function.
I will use the term “Orthogonal MUC” to the MUC.\end{thm}
\begin{proof}
Judging from the above\ref{thm:orthogonality and cyclic value assignment}\ref{thm:MUC as Completeness},
this is clear.
\end{proof}
Thus, by reducing logical value assignments to all cyclic value assignments
with keeping the orthogonality, we can express the MUC size in the
number of clauses.

\section{Orthogonalization of MUC}

We reduce MUC to orthogonal MUC under the constraints of the HornMUC.
First, we define the two inner product of clauses, and define the
orthogonality of clauses. Secondly, we clarify the limitations HornMUC.
And we show how to reduce the MUC to the orthogonal MUC.

And we show that we can reduce HornMUC to orthogonal MUC in polynomial
time from its connectivity, and can not reduce MUC.

I will use the term “Fact clause” to the HornCNF's clauses that include
only positive variable. {}``Goal clause'' to the HornCNF's clauses
that include only negative variable. {}``Rule clause'' to the HornCNF's
clauses that is not fact clause and goal clause. {}``Case clause''
to the HorncNF's clauses that is fact clauses and rule clauses.

\subsection{Clauses inner product and inner harmony}

Let us start with defining the inner product and inner harmony, and
considering the the these orthogonality. However, considering the
duality of the conjunction and disjunction, and consider the dual
inner product.
\begin{defn}
\label{def:Clause inner product}I will use the term “Inner product”
as follow;

$\left\langle C_{1},C_{2}\right\rangle =\left\langle C_{1}\bot C_{2}\right\rangle =\bigvee\left(C_{1}\left(x_{i}\right)\wedge C_{2}\left(x_{i}\right)\right)=\exists x_{i}\left(C_{1}\left(x_{i}\right)\wedge C_{2}\left(x_{i}\right)\right)$

\label{def:Clause inner harmony}I will use the term “Inner harmony”
as the duality of the inner product.

$\left\langle C_{1},C_{2}\right\rangle ^{d}=\left\langle C_{1}\top C_{2}\right\rangle =\overline{\bigvee\left(\overline{C_{1}\left(x_{i}\right)}\wedge\overline{C_{2}\left(x_{i}\right)}\right)}=\bigwedge\left(C_{1}\left(x_{i}\right)\vee C_{2}\left(x_{i}\right)\right)$

$=\neg\exists x_{i}\left(\overline{C_{1}\left(x_{i}\right)}\wedge\overline{C_{2}\left(x_{i}\right)}\right)=\forall x_{i}\left(C_{1}\left(x_{i}\right)\vee C_{2}\left(x_{i}\right)\right)$

$\bigvee\left(\right),\bigwedge\left(\right)$ is the disjunction
and conjunction of all truth value assignment value.

It also defines the inner product and inner harmony of more than two.
And it also defines the inner product and inner harmony of CNF. 
\end{defn}
Also, we can define the orthogonality of inner product and inner harmony.

depending on whether that is false in the inner section and can be
determined for each of the orthogonality. Defined as follows: Note
that the duality of inner product replace true and false.
\begin{defn}
\label{def:Inner product orthognality}When the inner product is false,
the clauses are orthogonal at inner product. I will use “$C_{1}\bot C_{2}$”
to orthogonal at inner product. Orthogonal CNF at inner product is
unsatisfiable CNF.

\label{def:Inner harmony orthognality}When the inner harmony is true,
the clauses are orthogonal at inner harmony. I will use “$C_{1}\top C_{2}$”
to orthogonal at inner harmony. Orthogonal CNF at inner harmony is
the CNF that all logical value assignment are cyclic value assignment.

\label{def:Inner harmony Orthogonalization}I will use term {}``Clauses
orthogonalization'' to reducing two clauses to orthogonal clauses
at inner harmony.
\end{defn}
When the inner harmony is orthognal, the CNF that include these clauses
are orthogonal at inner harmony. I will use “$C_{1}\top C_{2}$” to
orthogonal at inner harmony.

\subsection{HornMUC constraints}

Describes the HornMUC's constraints. HornMUC is the CNF that each
clauses have at most one positive variables. This restrict the phase
difference of clauses. Therefore, it is also restrict the inner harmony.

First, we show the restriction of the phase difference in the HornMUC
clauses.
\begin{thm}
\label{thm:HornMUC Connection}Phase difference between the two clauses
of HornMUC would be at most one. In other words, HornMUC clauses are
connected together.\end{thm}
\begin{proof}
We assume that HornMUC have two clauses $C_{1}=\left(x_{1}\vee\overline{x_{2}}\cdots\right),C_{2}=\left(x_{2}\vee\overline{x_{1}}\cdots\right)$.
These clauses are true when $T=\left(x_{1},x_{2}\cdots\right)=\left(\perp,\perp\cdots\right)$.
Thus, HornMUC must include the clause that is false at truth value
assignment $T$. But in order to satisfy this condition, HornCNF include
$\left(x_{1}\vee x_{2}\right)$ or be false regardless $\left(x_{1},x_{2}\right)$.
This is contradicts the assumption that HornMUC include $C_{1},C_{2}$.
Thus From the proof by contradiction, HornMUC do not include the clauses
these phase differnet more than two.
\end{proof}
We can see from the HornMUC restrict what the structure of HornCNF
is restricted. HornMUC's phase difference is at most one, HornMUC
structure has been connected not only at whole but also each part.
Thus, HornMUC can not construct the structure with the non-connected
part. HornMUC constitutes a partial order of phases. And clause cycle
do not affect to the HornMUC's partial order.
\begin{thm}
\label{thm:Connectivity constraints of HornMUC}HornMUC can not construct
the structure with the non-connected part. And HornMUC constitutes
a partial order of phases.\end{thm}
\begin{proof}
Shows the HornMUC's connection. From mentioned above \ref{thm:HornMUC Connection},
all clauses are connecting or crossing. We assume that there are HornMUC
that can be divided into two subsets of non-connected to each other.
We assume that we can split the HornMUC into two subsets that is not
connected. The subsets have no common variables or have only common
negative variables. But if the subsets have no common variables, HornMUC's
unsatisfiability do not change to delete one of these subset. So,
This is contradicts the assumption of HornMUC. And if the subsets
have only common negative variables, HornMUC's unsatisfiability do
not change to delete the clauses that include that negative variables.
So, from the proof by contradiction, we can not divide HornMUC into
two subsets of non-connected to each other.

Shows the HornMUC's structure of partially ordered. About HornCNF's
upper clause $C_{U}$ and lower clause $C_{L}$;

$\cdots\geq C_{U}\geq C_{k}\geq C_{k-1}\geq\cdots\geq C_{1}\geq C_{L}\geq\cdots$

$\rightleftarrows\cdots\geq\left(x_{U}\vee\overline{x_{k}}\cdots\right)\geq\left(x_{k}\vee\overline{x_{k-1}}\vee\cdots\right)$

$\geq\cdots\geq\left(x_{1}\vee\overline{x_{L}}\vee\cdots\right)\geq\left(x_{L}\vee\cdots\right)\geq\cdots$

It is clear that reflexive is satisfied. It is clear that transitivity
is satisfied from HornCNF's constraints (clause include at most one
positive variable.)

We can see the antisymmetric from the following;

$\forall C_{1},C_{2}\left(\left(C_{1}\leq C_{2}\right)\wedge\left(C_{2}\leq C_{1}\right)\right)$

$=\forall F_{i=1,2,3,4}\left(\forall x_{1},x_{2}\left(\left(x_{1}\vee F_{1}\right)\wedge\left(x_{2}\vee\overline{x_{1}}\vee F_{2}\right)\wedge\left(x_{2}\vee F_{3}\right)\wedge\left(x_{1}\vee\overline{x_{2}}\vee F_{4}\right)\right)\right)$

$=\forall x_{1},x_{2}\left(\left(x_{1}=x_{2}\right)\wedge\left(x_{2}\vee\overline{x_{1}}\right)\wedge\left(x_{1}\vee\overline{x_{2}}\right)\right)=\forall x_{1},x_{2}\left(x_{1}=x_{2}\right)$

$F_{i=1,2,3,4}:CNF$

So HornMUC clauses constitutes a partial order.
\end{proof}
This constrain HornMUC's inner harmony. HornMUC constitutes a partial
order from empty clause leading to fact clause, rule clause, goal
clause, and finally empty clause. Thus, we want to split the HornMUC,
we can only cut into upper and lower part of this partial order.
\begin{thm}
\label{thm:HornMUC cut}When we split the HornMUC, we can only cut
into upper and lower part of this partial order. Each part is a partial
order. Cutting clause exist only one part.\end{thm}
\begin{proof}
It is clear that HornMUC is a partial order.
\end{proof}
For example, if you divide the $\left(x_{0}\right)\wedge\left(\overline{x_{0}}\right)$
by MUC;

$\left(x_{0}\right)\wedge\left(\overline{x_{0}}\vee x_{1}\right)\wedge\left(\overline{x_{1}}\vee\overline{x_{2}}\vee\overline{x_{3}}\right)\wedge\left(\overline{x_{1}}\vee x_{2}\vee x_{3}\right)\wedge\left(x_{2}\vee\overline{x_{3}}\right)\wedge\left(\overline{x_{2}}\vee x_{3}\right)$

$\left(\overline{x_{0}}\vee x_{1}\right)$ is divide to $\left(\overline{x_{0}}\vee x_{1}\vee x_{2}\vee x_{3}\right)\wedge\left(\overline{x_{0}}\vee x_{1}\vee\overline{x_{2}}\vee\overline{x_{3}}\right)$.

But because HornMUC can not be greater than the distance of two clauses,
we can not divide any clauses.

\subsection{Clause Orthogonalization by using HornMUC}

Describes the way to Orthogonalize clause by using HornMUC. For mentioned
above \ref{thm:Connectivity constraints of HornMUC}\ref{thm:HornMUC cut},
we are constrained when split MUC by using HornMUC because of HornMUC's
partial order. And we must cut the clauses when we orthogonalize clauses
to orthogonal part and correlation part.

For example, if you orthogonalize $\left(x_{0}\vee\overline{x_{1}}\vee\overline{x_{2}}\right)$
and $\left(x_{0}\vee x_{3}\vee x_{4}\right)$, we must cut follows;

$\left(x_{0}\vee\overline{x_{1}}\vee\overline{x_{2}}\right)\wedge\left(x_{0}\vee x_{3}\vee x_{4}\right)$

$=\left(x_{0}\vee\overline{x_{1}}\vee\overline{x_{2}}\right)\wedge\left(x_{0}\vee x_{3}\vee x_{4}\vee x_{1}\right)\wedge\left(x_{0}\vee x_{3}\vee x_{4}\vee\overline{x_{1}}\right)$

$=\left(x_{0}\vee\overline{x_{1}}\vee\overline{x_{2}}\right)\wedge\left(x_{0}\vee x_{3}\vee x_{4}\vee x_{1}\right)$

$\wedge\left(x_{0}\vee x_{3}\vee x_{4}\vee\overline{x_{1}}\vee x_{2}\right)\wedge\left(x_{0}\vee x_{3}\vee x_{4}\vee\overline{x_{1}}\vee\overline{x_{2}}\right)$

$=\left(x_{0}\vee\overline{x_{1}}\vee\overline{x_{2}}\right)\wedge\left(x_{0}\vee x_{3}\vee x_{4}\vee x_{1}\right)\wedge\left(x_{0}\vee x_{3}\vee x_{4}\vee\overline{x_{1}}\vee x_{2}\right)$

repeat that cutting target clause to orthogonal part and correlation
part. And finally correlation clause is absorbed to another clauses
which match base clause. In addition, this reduction is added only
HornMUC, so we can keep CNF's satisfies possibility.
\begin{defn}
\label{thm:Clause cut}I will use term {}``Clause cut'' to divide
clause by using HornMUC.\end{defn}
\begin{thm}
\label{thm:Cut and Orthogonalization}By cutting correlation clauses
to orthogonal part and correlation part, we can orthogonalize the
clauses.\end{thm}
\begin{proof}
It is clear that we can achieved by generalizing the above procedure.
So I omit.
\end{proof}

\subsection{HornMUC Orthogonalization}

Describes that we can reduce HornMUC to orthogonalize MUC in polynomial
time. For mentioned above \ref{thm:Cut and Orthogonalization}, we
can reduce HornMUC to orthogonalize MUC by cutting all correlation
parts. And we can reduce more easier HornMUC to orthogonalize MUC
by using HornCNF's constraint. Specifically, we can reduce that we
cut the clause at the negative variable correspond to the positive
variable of lower clause that is nearer from fact clause. Thus, we
can absorb every cutting.

For example to Orthogonalize $\left(x_{0}\right)\wedge\left(\overline{x_{0}}\vee x_{1}\right)\wedge\left(\overline{x_{1}}\vee x_{2}\right)\wedge\left(\overline{x_{2}}\right)$,
cutting $\left(\overline{x_{1}}\vee x_{2}\right)$ at $x_{0}$, cutting
$\left(\overline{x_{2}}\right)$ at $x_{0}$, and cutting $\left(\overline{x_{0}}\vee\overline{x_{2}}\right)$
at $x_{1}$.

$\left(x_{0}\right)\wedge\left(\overline{x_{0}}\vee x_{1}\right)\wedge\left(\overline{x_{1}}\vee x_{2}\right)\wedge\left(\overline{x_{2}}\right)$

$=\left(x_{0}\right)\wedge\left(\overline{x_{0}}\vee x_{1}\right)\wedge\left(\overline{x_{0}}\vee\overline{x_{1}}\vee x_{2}\right)\wedge\left(x_{0}\vee\overline{x_{1}}\vee x_{2}\right)\wedge\left(\overline{x_{2}}\right)$

$=\left(x_{0}\right)\wedge\left(\overline{x_{0}}\vee x_{1}\right)\wedge\left(\overline{x_{0}}\vee\overline{x_{1}}\vee x_{2}\right)\wedge\left(\overline{x_{2}}\right)$

$=\left(x_{0}\right)\wedge\left(\overline{x_{0}}\vee x_{1}\right)\wedge\left(\overline{x_{0}}\vee\overline{x_{1}}\vee x_{2}\right)\wedge\left(\overline{x_{0}}\vee\overline{x_{2}}\right)\wedge\left(x_{0}\vee\overline{x_{2}}\right)$

$=\left(x_{0}\right)\wedge\left(\overline{x_{0}}\vee x_{1}\right)\wedge\left(\overline{x_{0}}\vee\overline{x_{1}}\vee x_{2}\right)\wedge\left(\overline{x_{0}}\vee\overline{x_{2}}\right)$

$=\left(x_{0}\right)\wedge\left(\overline{x_{0}}\vee x_{1}\right)\wedge\left(\overline{x_{0}}\vee\overline{x_{1}}\vee x_{2}\right)\wedge\left(\overline{x_{0}}\vee\overline{x_{1}}\vee\overline{x_{2}}\right)\wedge\left(\overline{x_{0}}\vee x_{1}\vee\overline{x_{2}}\right)$

$=\left(x_{0}\right)\wedge\left(\overline{x_{0}}\vee x_{1}\right)\wedge\left(\overline{x_{0}}\vee\overline{x_{1}}\vee x_{2}\right)\wedge\left(\overline{x_{0}}\vee\overline{x_{1}}\vee\overline{x_{2}}\right)$

Because the number of clauses do not increase and the number of variables
in each clauses increase at most polynomial size, so this reduction
increase at most polynomial size. And HornMUC clauses constitutes
a partial order and it is enough to orthogonalize each clause at lower
clauses. So the reduction HornMUC to orthogonal MUC takes at most
polynomial time.
\begin{thm}
\label{thm:HornMUC orthogonality}We can reduce HornMUC to orthogonal
MUC at most polynomial size and time. \end{thm}
\begin{proof}
It is clear that we can achieved by generalizing the above procedure.
So I omit.\end{proof}
\begin{thm}
\label{thm:Total order of orthogonalized HornMUC}The clauses and
variables of the orthogonal MUC that reduced from Horn MUC constitute
total order.\end{thm}
\begin{proof}
From the above procedure, the clause of orthogonal MUC include the
negative variable that's positive variable is included in the lower
clauses. So, all clauses have order, and orthogonal MUC constitute
total order.\end{proof}
\begin{thm}
\label{thm:MUC orthogonalization and computation complexity}If we
can not reduce MUC to orthogonal MUC in polynomial time by using clause
cutting, then $P\neq NP$.\end{thm}
\begin{proof}
From the above \ref{thm:Total order of orthogonalized HornMUC}, if
MUC is HornMUC, then we can reduce the MUC to orthogonal MUC by using
clause cutting in polynomial time and size. And if $P=NP$, then we
can reduce MUC to HornMUC in polynomial time and size. So, if $P=NP$
and we can reduce MUC to HornMUC in polynomial time, then we can reduce
the MUC to orthogonal MUC in polynomial time and size.

Taking the contrapositive, if we can not reduce MUC to orthogonal
MUC in polynomial time and size, then $P\neq NP$ or we can not reduce
MUC to HornMUC in polynomial time and size. So, from the above \ref{thm:Relation between DP and P},
if we can not reduce MUC to orthogonal MUC in polynomial time and
size by using clause cutting, then $P\neq NP$.
\end{proof}

\subsection{MUC Orthogonalization}

Describes that we can not reduce MUC to orthogonalize MUC in polynomial
time. Unlike HornMUC, MUC can be set to any phase difference between
the clauses. So MUC clauses have high dispartion and correlation.
But for mentioned above \ref{thm:Connectivity constraints of HornMUC},
HornMUC connected each clauses. So we must use many HornMUC each other
to cut every dispart areas to orthogonal part and correlation part.
And because of HornMUC's connection, we can not use same HornMUC to
cut different areas.

For example to this, we think MUC include follow clauses;

$O_{3}\left(x_{0},x_{1},x_{2}\right)\wedge E_{3}\left(x_{3},x_{4},x_{5}\right)$

$O_{3}\left(x_{0},x_{1},x_{2}\right)=\left(x_{0}\vee x_{1}\vee x_{2}\right)\wedge\left(x_{0}\vee\overline{x_{1}}\vee\overline{x_{2}}\right)\wedge\left(\overline{x_{0}}\vee x_{1}\vee\overline{x_{2}}\right)\wedge\left(\overline{x_{0}}\vee\overline{x_{1}}\vee x_{2}\right)$

$E_{3}\left(x_{3},x_{4},x_{5}\right)=\left(\overline{x_{3}}\vee\overline{x_{4}}\vee\overline{x_{5}}\right)\wedge\left(\overline{x_{3}}\vee x_{4}\vee x_{5}\right)\wedge\left(x_{3}\vee\overline{x_{4}}\vee x_{5}\right)\wedge\left(x_{3}\vee x_{4}\vee\overline{x_{5}}\right)$

In other words, $O_{3}$ is true when it contains an odd trues in
truth value assignment, $E_{3}$ is true when it contains even trues
in truth value assignment. $O_{3}$ and $E_{3}$ divide the truth
value assignment. So if we want to orthogonalize the MUC that include
$O_{3}$ and $E_{3}$, we must cut every area by using every other
HornCNF. This orthogonalize MUC is the MUC that expanded to CNF;

$O_{3}\left(x_{0},x_{1},x_{2}\right)$

$\wedge\left(E_{3}\left(x_{3},x_{4},x_{5}\right)\vee\left(\left(\overline{x_{0}}\vee\overline{x_{1}}\vee\overline{x_{2}}\right)\wedge\left(\overline{x_{0}}\vee x_{1}\vee x_{2}\right)\wedge\left(x_{0}\vee\overline{x_{1}}\vee x_{2}\right)\wedge\left(x_{0}\vee x_{1}\vee\overline{x_{2}}\right)\right)\right.$
\begin{thm}
\label{thm:Limitation of orthogonalization by using HornMUC}When
we orthogonalize CNF that disparted some area, We must cut each area
by using every another HornMUC.\end{thm}
\begin{proof}
To assume that we can orthogonalize two area by using one HornMUC.
This time, there is other closure area between the divided area. From
assuming, we can orthogonalize these area by using one HornMUC. But
we must cross the HornMUC at the closure that divided area. This is
contradicts the assumption that we can orthogonalize by using one
HornMUC. Thus, From the proof by contradiction, we can not orthogonalize
two area by using one HornMUC.
\end{proof}
As a example, let us construct the MUC that have many divided area
amount of non-polynomial size of MUC. We should notice that the truth
value assignments having same even-odd of true value do not connect
each other. 

By combining proper $O_{3}$ and $E_{3}$ above, it is possible to
configure the CNF that is true when the truth value assignment having
same even-odd of true. For example;

$O_{4}\left(x_{0},x_{1},x_{2},x_{3}\right)=O_{3}\left(x,x_{0},x_{1}\right)\wedge E_{3}\left(x,x_{2},x_{3}\right)$

This CNF is true only if the truth value having odd of true value.

This can be extended easily for any number of variables.

$O_{n+1}\left(x_{0},\cdots,x_{n}\right)=O_{n}\left(x,x_{0},\cdots,x_{n-2}\right)\wedge E_{3}\left(x,x_{n-1},x_{n}\right)$

Nunber of $O_{n}$ clause is polynomial size of variables. And truth
value assignment having odd of true value is amount of half of all
truth value assignment, and every truth value assignment having same
even-odd of true do not connect each other. So $O_{n}$ have many
divided area amount of non-polynomial size. And there is MUC that
we can not reduce to orthogonal MUC in polynomial size.
\begin{thm}
\label{thm:size of orthogonal MUC}there is MUC that we can not reduce
to orthogonal MUC in polynomial size by using HornMUC.\end{thm}
\begin{proof}
For mentioned above examples, there is CNF that divide the truth value
assignment into the same even-odd of true value. And these CNF can
be a part of the MUC. So there is the MUC that divided area amount
of non-polynomial size. For mentioned above \ref{thm:Limitation of orthogonalization by using HornMUC},
we must cut each area to orthogonalize MUC. So there is MUC that we
can not reduce to orthogonal MUC in polynomial size.
\end{proof}

\section{DP is not P and NP is not P}

Describes the difference between MUC decision problem and HornMUC
decision problem. For mentioned above \ref{thm:1.3 CNF classification and MUC dicision problem},
the difference of the MUC decision problem and HornMUC decision problem
will appear in CNF classification. And orthogonal basis of inner harmony
is different between MUC and HornMUC. For mentiond above \ref{thm:HornMUC orthogonality},
all HornMUC have polynomial size orthogonal basis. But For mentiond
above \ref{thm:size of orthogonal MUC}, there is MUC that have non-polynomial
size orthogonal basis. All orthogonal MUC's clause is orthogonal basis,
and correspond to the equivalence class. So we can not reduce MUC
to HornMUC in polynomial size. So $DP\neq P$. And for mentiond above
\ref{thm:MUC orthogonalization and computation complexity}, $NP\neq P$.


\begin{thebibliography}{2}
\appendix
\bibitem{Symmetry and Uncountability of Computation}Koji~KOBAYASHI,~Symmetry~and~Uncountability~of~Computation.~2010,~2010arXiv1008.2247K

\bibitem{Computational Complexity}C.~H.~Papadimitriou~and~M.~Yannakakis.~The~complexity~of~facets~(and~some~facets~of~complexity),~Journal~of~Computer~and~System~Sciences~28:244-259,~1984. 
\end{thebibliography}
\end{document}